\documentclass[aps,twocolumn,nofootinbib]{revtex4}
\usepackage[latin1]{inputenc}
\usepackage{epsfig}
\newcommand{\beq}{\begin{equation}}
\newcommand{\eeq}{\end{equation}}
\newcommand{\la}{\langle}
\newcommand{\ra}{\rangle}

\begin{document}

\title{Stochastic thermodynamics
of ecosystems}

\author{Tânia Tomé and Mário J. de Oliveira}
\affiliation{Universidade de São Paulo, Instituto de Física,
Rua do Matão, 1371, 05508-090 São Paulo, SP, Brazil}

\begin{abstract}

We investigate the thermodynamics as well as the population
dynamics of ecosystems based on a stochastic approach
in which the number of individuals of the several species
of the ecosystem are treated as stochastic variables.
The several species are connected by feeding relationships
that are understood as unidirectional processes in which 
a certain amount of biomass is exchanged between species.
We show that the equations for the averages in the numbers of
individuals are that given by the deterministic approach.
We determine the fluxes of mass, energy, and entropy as well
as the rate of the entropy production.
This last quantity, which has a central role in the present
stochastic approach, is obtained by a formula appropriate for
unidirectional transitions. The flux of energy across the
ecosystem is shown to be proportional to the flux of entropy
to the environment.

\end{abstract}

\maketitle

\section{Introduction}

An ecosystem consists of organisms that live in the same
space, interacting with each other and with the physical
environment 
\cite{warming1909,tansley1935,elton1927,odum1953,macarthur1955,
odum1983,ricklefs2000,ricklefs2008}. 
The organisms of distinct species are connected with each
other through a feeding relationship structured in
a hierarchy of trophic levels, called food web
\cite{elton1927,odum1953,macarthur1955,odum1983,
ricklefs2000,ricklefs2008,paine1966}. 
In the first trophic
level one finds the organisms that produce organic matter
from inorganic substances. These are the autotrophs.
In the other levels we find the heterotrophs which obtain
organic matter by feeding on the autotrophs and on other
heterotrophs. In the second level, there are
the species that eat the autotrophs and are food for the
species of the third level. The species of this level are
in turn food for the upper level and so on. The top level
consists of species that are not food for any other species
and are represented by the apex predators.

The feeding relationship induces a change in the number of
individuals of each species which for that reason evolves in
time and may reach a stationary state. Many approaches have
been employed in the theoretical study of the sizes of
populations in food webs and its evolution in time
\cite{may1973,smith1974,pimm1982,deruiter2005}.
We point out the deterministic approach in which the
number of individuals of each species obeys an ordinary
differential equation of first order in time. This
approach was employed by Lotka \cite{lotka1922,lotka1925} and
by Volterra \cite{volterra1931} in their study of a
predator-prey system. 
The Lotka-Volterra model was extended to several
interacting species by May in his studies of the stability
in mutispecies community models \cite{may1971}.
The extended model was then used by other investigators
as a model for food webs 
\cite{pimm1979,pimm1987,moore1993,deruiter1995}.

The structure composed of the organisms and
the abiotic environment is maintained active
by the consumption
of light energy by the autotrophs that transform inorganic
substances into organic matter through photosynthesis.
The heterotrophs obtain organic matter by feeding on
the autotrophs and on other heterotrophs. These organisms
convert the nutrients into matter that are used again by
the autotrophs completing the cyclic transformation of matter. 
The transformation of matter induced by the input
of energy in an ecosystem were pointed out
by Lotka \cite{lotka1922,lotka1925} and by Lindeman
\cite{lindeman1942}, 
and the role of the flow of energy trough the system
was emphasized by Odum \cite{odum1953,odum1983}. 
The flow of energy acts to organize the system
\cite{morowitz1968}, 
and an ecosystem is regarded as a thermodynamic system
which transforms matter and maintain the living 
structure through the flow of energy.

The continuous flow of energy through the system
shows that the equilibrium thermodynamics cannot
be applied because in thermodynamic equilibrium
there can be no macroscopic flow of any kind and
particularly of energy. Therefore, it is necessary
to resort to theories that take into account
the irreversible character of the processes such
as that developed by De Donder and by Prigogine
\cite{dedonder1927,prigogine1946,prigogine1947,
prigogine1955,glansdorff1971,nicolis1977} 
in which entropy production, entropy flux and energy flux
are central concepts. Thus nonequilibrium
thermodynamics has been applied to ecosystems
\cite{zotin1967,feistel1981,assimacopoulos1986,auger1989,
chakrabarti1995,svirezhev2000,jorgensen2004,michaelian2005,
chakrabarti2009,andrae2010,nielsen2020}
We wish here to apply the stochastic thermodynamics
to ecosystem and thus explains through this theory the
flows of matter and energy and their connection with the
flow of entropy and the production of entropy.
The stochastic thermodynamics
\cite{tome2010,tome2015,tome2018,tome2023a}
is based on a stochastic dynamics for the
time evolution of the system. Like statistical mechanics
the states of the system are defined by a set of
random variables over which a probability distribution is
defined. The stochastic dynamics is represented by a master
equation or by a Fokker-Planck equation.

The stochastic dynamics are defined by the transition
rates associated to the several processes occurring
within the system.
From the transition rates one obtains not only the 
equations that governs the time evolution of the
probability distribution but also the production of entropy.
The formula for the production of entropy that we
use here is distinct from the usual expression
introduced by Schnakenberg 
\cite{schnakenberg1976}. This expression
is appropriate for transitions that have the reverse
but breaks down for unidirectional transitions, which
are the case of feeding process treated here.
Thus we propose here a formula which is appropriate for
unidirectional transitions.
In the following, we use the deterministic approach
to population dynamics of ecosystem, and then proceed
to present the stochastic approach, including the
concepts of production of
entropy, flux of entropy and energy flux.

\section{Deterministic dynamics}

\subsection{Dynamic equations}

An example of a food web is that consisting of plants, herbivorous
and carnivorous living in a given region. The first trophic level
of the food web consists of plants which generate organic
matter using light energy through photosynthesis. In the
second trophic level we find the herbivorous which eat plants and
are the food for the first carnivorous. The herbivorous as well as
the first carnivorous are the food for the second carnivorous.
There might be other carnivorous trophic levels until the
top level of the food web consisting of the apex carnivorous.

\begin{figure}
\epsfig{file=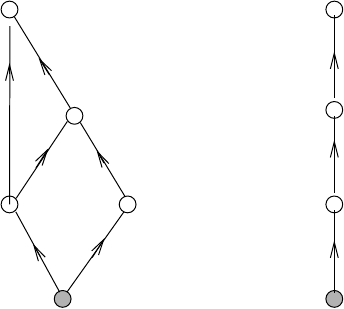,width=5cm}
\caption{Representation of two food webs with four trophic
levels. The gray circle represents the plants and the other
circles represent the animal species. An arrow represents the
process of feeding and points from an organism eaten to 
the animal eating it.}
\label{duasredes}
\end{figure}

A food web is represented by a set of nodes and
connections between them as shown in figure \ref{duasredes}. 
Each node of the food web represents an animal species
except the bottom node which represents the plants.
The feeding connection between the nodes are represented
by an arrow which points from an organism eaten to the
animal eating it. Several arrows may point to a node
meaning that the species at the node eats several species.
A node may also be at the tail of several arrows
meaning that the species is the food of several species.

Instead of using the numbers of individual of each species
as the dynamic variables as is usual in population dynamics,
we will employ dynamics variables that are related the biomasses
of each species. We introduce a standard mass $m$ and define
the dynamic variable $N_i$ associated to the species $i$ by
$N_i=M_i/m$ where $M_i$ is the biomass of species $i$, which
is the sum of the masses of the individuals of species $i$.
The biomass of the plants is denoted by $M_0$, and
$N_0=M_0/m$.

The biomasses vary in time according to the feeding processes
represented by the arrows in figure \ref{duasredes}. 
In this process a certain amount of the biomass of a
species is transferred to the biomass of the species
eaten it. In addition to this type of process there
is another type involving only the animals which is
the spontaneous annihilation of the biomass. 
A third type of process involves only the plants                   
and corresponds to transformation of inorganic substances
into organic matter trough photosynthesis.

To set up the equation that gives the time variation
of $N_i$, we consider first the contribution
coming from the feeding process. We assume that this
process is analogous to an autocatalytic reaction which
means that the contribution associated to species
$i$ and $j$ is proportional to the product $N_is_j$,
where $s_j$ is the fraction in mass of specie $j$.  
The contribution coming from the spontaneous annihilation of
the biomass of species $i$ is proportional to $N_i$.
Adding these two contributions we get the following
equation
\beq
\frac{dN_i}{dt} = \sum_{j=0}^n \nu_{ij} b_{ij} N_i s_j
-c_i N_i,
\label{3}
\eeq
where $b_{ij}$ is the feeding rate constant, $c_i$ is the
annihilation rate constant, and $\nu_{ij}=+1$ if $i$ feeds on $j$,
and $\nu_{ij}=-1$ otherwise. We remark that $c_i$ cannot be
zero for the apex carnivorous otherwise its number would
increase without bounds.

\begin{figure*}
\epsfig{file=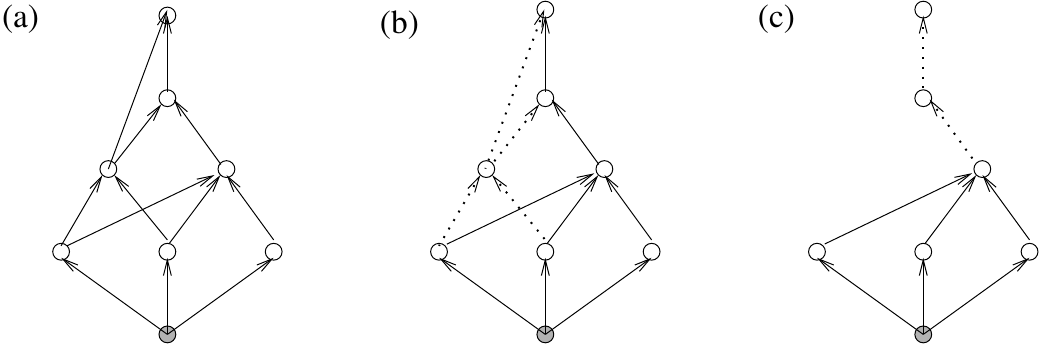,width=14cm}
\caption{(a) Representation of a food web with five trophic levels.
(b) Representation after one of the species
of the third level becomes extinct.
(c) Representation after the species of the fourth level
becomes extinct. As a consequence, the apex predator
of the fifth level also becomes extinct by the suppression
of its prey.}
\label{foodweb}
\end{figure*}

The third process, which involves only the plants, is represented
by a constant $Q$ which is proportional to the rate of production of organic
matter by the plants. The equation that gives the time variation
of $N_0$ is
\beq
\frac{dN_0}{dt} = - \sum_{j=1} b_{0j} N_j + Q.
\label{3a}
\eeq

It is more convenient to write the dynamic equations in terms
of the mass fraction $s_i$ defined as the ratio $s_i=N_i/N$ between
$N_i$ and $N$, where $N$ is a fixed quantity which is a measure
of the population size.
Dividing equations (\ref{3}) and (\ref{3a}) by $N$, we find
\beq
\frac{ds_i}{dt} = \sum_{j=0}^n \nu_{ij}b_{ij} s_is_j
- c_i s_i,
\label{9}
\eeq
valid for $i=1,2,\ldots,n$, and
\beq
\frac{ds_0}{dt} = - \sum_{j=1}^n b_{0j} s_0s_j + q,
\label{4}
\eeq 
where $q=Q/N$.

There are some immediate consequences
of the dynamic equations that are worth mentioning,
which are illustrated in figure \ref{foodweb}. 
If the biomass fraction $s_k$ of species $k$
vanishes then it remains zero forever,
which means that the species becomes extinct. Indeed,
if $s_k=0$, then it follows from equation
(\ref{9}) that $ds_k/dt=0$ and $s_k$ will
not change from its zero value. 

If one animal species becomes extinct
then the food web that remains is the one we
obtain by removing all the arrows that are connected
to it, as shown in figure \ref{foodweb}. Indeed,
if $k$ is the species going extinct then $s_k=0$.
Setting $s_k=0$ in the other equations is
equivalent to set $b_{ik}=0$, which corresponding
to erasing the corresponding arrow.

If all prey of a species $k$ disapear then
the species $k$ becomes extinct.
In other words, if all arrows pointing to $k$
are erased then $s_k=0$. Indeed, 
if all prey of $k$ disappears then
all positive terms on the right-hand
side of the equation (\ref{9})
disappear, remaining only negative terms, so that
$ds_k/dt<0$ and $s_k$ eventually vanishes.

\subsection{Food chains}
\label{foodchains}

We consider here food chains which are food webs
such that each animal has just one prey 
like that shown on the right panel of figure \ref{duasredes}.
The species $i+1$ feeds on the species $i$.
To turn the model simpler we set $b_{i+1,i}=b$,
except $b_{10}=a$, and $c_i=c$.
The equations become
\begin{eqnarray}
\frac{ds_1}{dt} &=& a s_1s_0  - b s_1s_2- c s_1, \\
\frac{ds_i}{dt} &=& b s_is_{i-1}  - b s_is_{i+1}- c s_i, \\
\frac{ds_n}{dt} &=& b s_ns_{n-1} - c s_n, \\
\frac{ds_0}{dt} &=& - a s_0s_1 + q.
\end{eqnarray}

The nonzero stationary solution is determined by the
set of equations  
\begin{eqnarray}
a s_0 - bs_2 &=& c, \\
b s_{i-1}  - b s_{i+1} &=&  c, \\
b s_{n-1} &=& c, \\
a s_0s_1  &=& q.
\end{eqnarray}

\begin{figure}
\epsfig{file=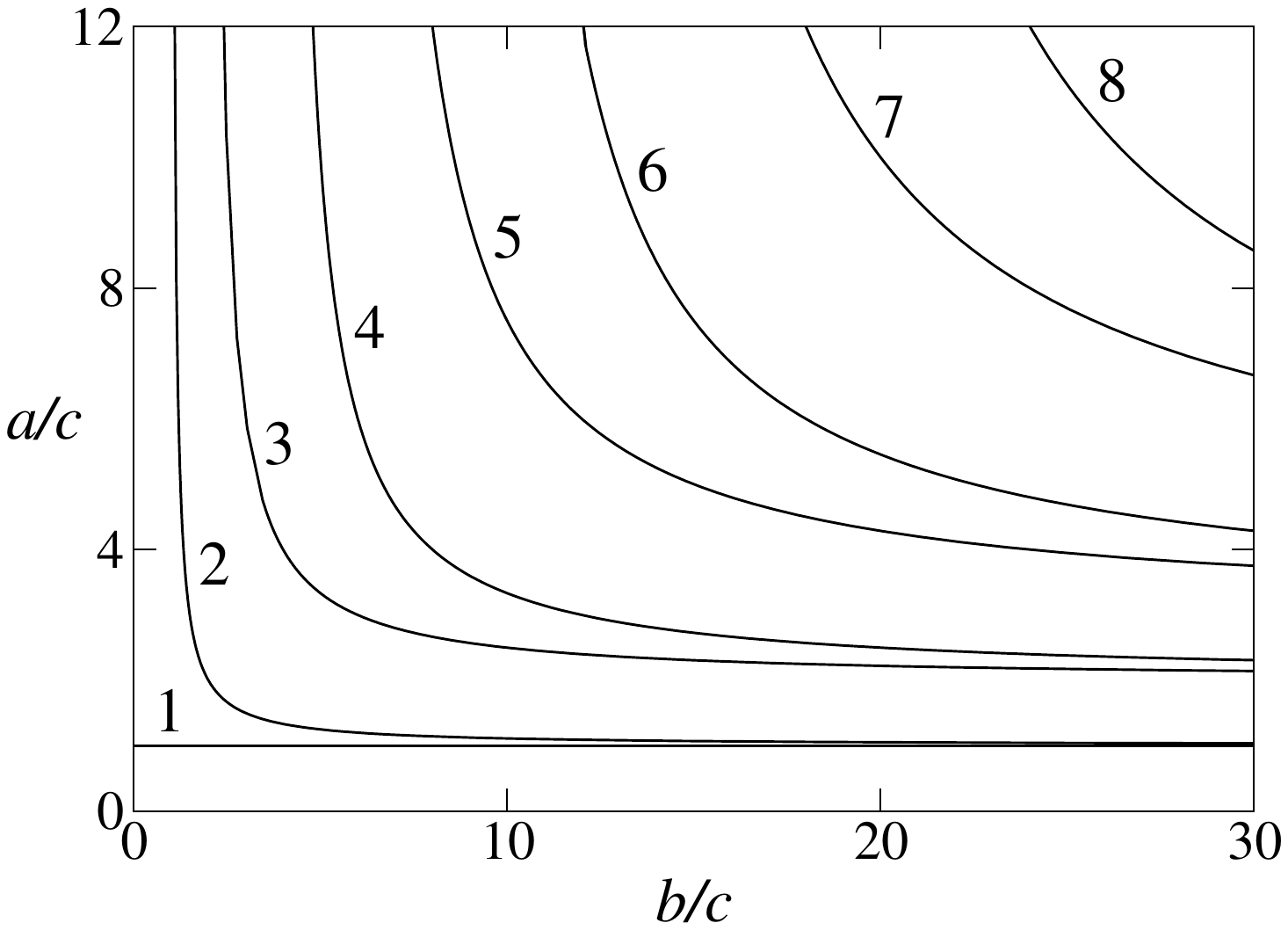,width=7.4cm}
\caption{Diagram in the space of parameters
$a/c$ versus $b/c$ for the food chain with $n$
animal species. The number next to a line
corresponds to the number $n$ of animal species.
A line $n$ determines the boundary of the
region ${\cal R}_n$ of the occurrence of the positive solution
$s_i>0$, for $i=0,1,2,\ldots,n$, which occurs
at the right and above the line $n$.}
\label{traline}
\end{figure}

It is straightforward to solve this system of
equations. For $n$ even the solution is
\begin{eqnarray}
s_0 &=& \frac{2b q}{nac}\\
s_\ell &=& (n-\ell+1)\frac{c}{2b}, 
\qquad \ell \,\,{\rm odd}, \\
s_\ell &=& \frac{2q}{nc} - \frac{\ell c}{2b},
\qquad \ell \,\,{\rm even}.
\end{eqnarray}
The fractions $s_0$ and $s_\ell$ for $\ell$ odd are all
positive. The smaller fraction $s_\ell$ 
for $\ell$ even is $s_n$. Thus the condition
that all $s_\ell$ be positive is given by $s_n>0$, or
\beq
\frac{2q}{nc} > \frac{nc}{2b},
\eeq
which defines the region ${\cal R}_n$ in the space of
parameters where the full positive solution exists.
The boundary of this region is given by
\beq
\frac{2q}{nc} = \frac{nc}{2b},
\label{25a}
\eeq
valid for $n=2,4,6,\ldots$ and is shown in
figure \ref{traline} for the value of $q$
such that $s_0+s_1+\ldots+s_n=1$. 

Let us consider now the case $n$ odd.
In this case the solution of the system of
linear equation is
\begin{eqnarray}
s_0 &=& \frac{(n+1)c}{2a} \\
s_\ell &=& \frac{2q}{(n+1)c} -\frac{(\ell-1)c}{2b},
\qquad \ell \,\,{\rm odd}, \\
s_\ell &=& (n+1-\ell)\frac{c}{2b},
\qquad \ell \,\, {\rm even}.
\end{eqnarray}

The fractions $s_\ell$ for $\ell$ even are all positive.
The smaller fraction $s_\ell$ for $\ell$ odd is $s_n$.
Thus again the condition that all $s_\ell$ be positive
is given by $s_n>0$, or
\beq
\frac{2q}{(n+1)c} > \frac{(n-1)c}{2b},
\eeq
which defines the region ${\cal R}_n$
where the full positive solution exists.
The boundary of this region is given by
\beq
\frac{2q}{(n+1)c} = \frac{(n-1)c}{2b}, 
\label{25b}
\eeq
valid for $n=1,3,5,\ldots$ and is shown in
figure \ref{traline} for the value of $q$
such that $s_0+s_1+\ldots+s_n=1$.

It is worth mentioning that as one approaches
the boundary of ${\cal R}_n$, given by 
(\ref{25a}) for $n$ even and by (\ref{25b}) for
$n$ odd, the fraction that vanishes is
that corresponding to the apex predator,
which becomes extinct at the boundary of
${\cal R}_n$.

\section{Stochastic dynamics}

\subsection{Fokker-Planck equation}

The dynamic approach that we have presented above describes
a deterministic motion. In the following we present a stochastic
approach to the thermodynamic of an ecosystem. The mass fraction
now becomes a stochastic variable, which we denote by $x_i$.
We use the notation $x$ to represent the collection of all
fractions $x_i$.
We begin by setting up a master equation that gives the
the evolution equation of the probability distribution of
$P(x)$ of the mass fractions. 

\begin{figure}
\epsfig{file=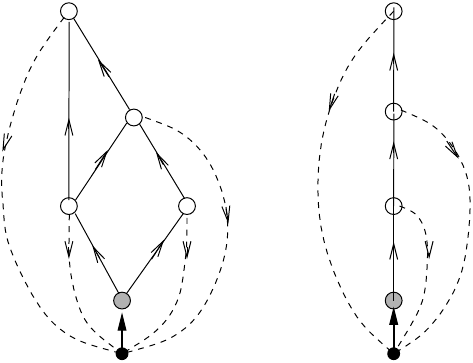,width=7cm}
\caption{
Representation of the processes of transfer of masses ocurring
in a ecosystem. The black small circle represents the surroundigs
containing inorganic matter. The gray circle represents
the plants and the other circles represent the animal species.
The solid arrow at the botton pointing from the black to the gray
circle represents the process of production of organic matter
by the plants through photosyntesis. 
The other solid arrows represent the process
of feeding. A solid arrow points from an organism eaten to 
the animal eating it. The dashed arrows represent
the spontaneous annihilation of animals and
the transformation of organic into inorganic matter.}
\label{twowebs}
\end{figure}

The ecosystem to be described here by the master equation 
is understood as composed by $n+2$ components,
as shown in figure \ref{twowebs}, each one labeled by an integer
$i$ from 0 to $n+1$. The component $i=n+1$ represents the
surroundings consisting of inorganic matter. The component $i=0$
represents the plants, that transform inorganic matter into organic
matter. The components from $i=1$ to $i=n$ represents each one an
animal species, that eats another animal or the plants.

We remark that the mass of the surrounding consists
of inorganic matter. The plants transforms the inorganic
matter into the organic matter, a process represented by the
lowest vertical arrow in figure \ref{twowebs}. The  
solid arrows represent the 
process of feeding which transfers biomass from the
species being eaten to the eating species.
The processes represented by dashed arrows correspond
to the annihilation of animals and are understood as
a transformation of organic to inorganic matter
deposited in the surroundings.

All these processes conserve mass and correspond to the transfer
of a certain mass from one component $j$ to another $i$. The mass
transferred is taken to be equal to $m$, the standard mass
introduced above. This is equivalent to say that a biomass $M_i$
increases or decreases by and amount $m$ or that the variable
$N_i=M_i/m$ increases or decreases by unity. Thus the mass fraction
increases or decreases by an amount $\varepsilon=1/N$. A processes
represented by an arrow in figure \ref{twowebs} from $j$ to $i$  
corres corresponds to a decrease of the fraction
$x_j$ by an amount $\varepsilon$ and an increase of the fraction $x_i$ 
by the same amount. We represent this process by the transition
\beq
x \to x^{ij},
\eeq
where $x$ represents the vector 
\beq
x = (x_0,\ldots,x_i,\ldots,x_j,\ldots,x_{n+1}),
\eeq
and $x^{ij}$ is the vector
\beq
x^{ij} = (x_0,\ldots,x_i+\varepsilon,\ldots,
x_j-\varepsilon,\ldots,x_{n+1}),
\eeq
obtained from $x$ by replacing $x_i$ by $x_i+\varepsilon$
and $x_j$ by $x_j-\varepsilon$.

The transition $x \to x^{ij}$ occurs with probability per unit which
we denote by $w_{ij}(x)$ and may depend on $x$. It is given by
\beq
w_{ij} = b_{ij} x_ix_j,
\label{31a}
\eeq
if $i=1,\ldots,n$ and $j=0,1,\ldots,n$, where $b_{ij}$ is
the feeding rate constant.
When $i=n+1$ and $j=1,\ldots,n$ then
\beq
w_{n+1,j} = c_j x_j,
\label{31b}
\eeq
where $c_j$ is an annihilation rate constant.
When $i=0$ and $j=n+1$ then
\beq
w_{0,n+1} = q,
\label{31c}
\eeq 
where $q$ is the rate in which the organic matter is created
by the plants.

Once we are given the transition rates, the master equation
reads
\beq
\frac{d}{dt}P(x) = \frac1\varepsilon \sum_{ij}
\{w_{ij}(x^{ji}) P(x^{ji}) - w_{ij}(x)P(x)\},
\label{28}
\eeq
where $i$ and $j$ take all values from 0 to $n+1$.
As $\varepsilon$ is small compared to $x_i$ we may expand the
right hand side of the master equation in powers of $\varepsilon$.
Up to linear terms in $\varepsilon$, we find
\beq
\frac{dP}{dt} = \sum_{ij}
\{- D_{ij}w_{ij} P + \frac{\varepsilon}2 D_{ij}^2 w_{ij} P\},
\label{23}
\eeq
where $D_{ij}=\partial/\partial x_i-\partial/\partial x_j$,
which is a Fokker-Planck equation. The presence of the quantity
$\varepsilon$ in the last term of the Fokker-Planck indicates
that $\varepsilon$ is a measure of the fluctuations, that is,
the covariances of $x_i$ are proportional to $\varepsilon=1/N$.
In a equivalent manner we may say that the amplitudes of
the fluctuations in $N_i$ is proportional to $\sqrt{N}$.

From equation (\ref{23}) we may determine the
time evolution of the average $\la x_i\ra$. It is given by
\beq
\frac{d}{dt}\la x_i\ra =  \sum_j \la w_{ij}- w_{ji}\ra,
\eeq
where we have performed an integral by parts.
Taking into account that the probability distribution is
very sharped at the mean values $s_i=\la x_i\ra$, we
may replace $\la w_{ij}(x)\ra$ by $w_{ij}(s)$ and write
\beq
\frac{ds_i}{dt} =  \sum_j \{ w_{ij}(s)- w_{ji}(s)\}.
\label{26}
\eeq
We remark that this is a close set of equations for
$s_i$ and determine $s_i$ as a function of time.
This equation coincides with the deterministic
equations (\ref{9}) and (\ref{4}), and we may say that
within the present stochastic aproach, the dynamic
variables of the deterministic approach can be understood
as the average values of the stochastic variables.

\subsection{Entropy flux}

The entropy $S$ of the ecosystem is defined by
\beq
S = - k\int P\ln P dx,
\eeq
where $k$ is the Boltzmann constant and
the integration is performed over the space
$x=(x_0,x_1,x_2,\ldots,x_n,x_{n+1})$. Its time evolution is
\beq
\frac{dS}{dt} = - k\int \frac{\partial P}{\partial t} \ln P dx.
\eeq
Using the master equation (\ref{28}), we find
\beq
\frac{dS}{dt} = \frac{k}\varepsilon \sum_{ij}\int 
w_{ij}(x) P(x) \ln \frac{P(x)}{P(x^{ij})} dx.
\eeq

To determine the rate of entropy production we
propose the following formula 
\beq
\Pi = \frac{k}\varepsilon \sum_{ij}\int w_{ij}(x) 
\{P(x) \ln \frac{P(x)}{P(x^{ij})} - P(x) + P(x^{ij})\} dx,
\eeq
and we remark that $\Pi$ is nonnegative because if we let
$\xi=P(x)/P(x^{ij})$ we see that the integrand is
proportional to $\xi\ln\xi-\xi+1\geq0$.
This formula for the entropy production rate is distinct
from the usual Schnakenberg formula \cite{schnakenberg1976},
and is appropriate
for the unidirectional transitions that we consider here.

The time variation of the entropy of the system is equal to
the production of entropy minus the flux of entropy {\it from}
the system {\it to} the outside $\Psi$, that is,
\beq
\frac{dS}{dt} = \Pi - \Psi.
\eeq
It is worth mentioning that 
some authors call $dS/dt$, $\Pi$, and $\Psi$, respectively,
the system entropy production, denoted by $\dot{S}_{\rm sys}$,
the total entropy production, denoted by $\dot{S}_{\rm tot}$,
and the environment entropy production, denoted by $\dot{S}_{\rm env}$
\cite{busiello2020}.

From the expressions of $dS/dt$ and $\Pi$ we find the following
expression for the flux of entropy,
\beq
\Psi = \frac{k}\varepsilon \sum_{ij}\int w_{ij}(x) 
\{P(x^{ij}) - P(x)\} dx.
\eeq
It can be expressed as an average 
\beq
\Psi = \frac{k}\varepsilon \sum_{ij}\la w_{ij}(x^{ji}) - w_{ij}(x)\ra,
\eeq
and in the limit $\varepsilon\to0$, it becomes
\beq
\Psi = - k\sum_{ij} \la D_{ij} w_{ij}\ra.
\eeq
It is convenient to write $\Psi$ as a sum of terms
\beq
\Psi = \sum_{ij} \Psi_{ij},
\eeq
where
\beq
\Psi_{ij} = - k\la D_{ij} w_{ij}\ra,
\label{32}
\eeq
which is nonzero only when there is an arrow from $j$ to $i$.
In other cases $\Psi_{ij}$ vanishes.

Using the expressions (\ref{31a}), (\ref{31b}), and (\ref{31c}),
we find, respectively,
\beq
\Psi_{ij} = - k b_{ij}(s_j-s_i) \qquad i=1,2\ldots,n,
\label{33a}
\eeq
for the first type of transition,
\beq
\Psi_{n+1,j} = -k c_j,
\label{33b}
\eeq
for the second type, and
\beq
\Psi_{0,n+1}=0,
\label{33c}
\eeq
for the third type, which means that there is no flux of entropy
associated to the input of energy through the plants.

\subsection{Energy flux}

Let ${\cal E}(x)$ be a state function, which we wish to identify 
with the energy function. The time variation of its
average $U=\la {\cal E}\ra$ is
\beq
\frac{dU}{dt} = \int \frac{\partial P}{\partial t} {\cal E} dx.
\eeq
Using the master equation (\ref{28}) we find
\beq
\frac{dU}{dt} = \Phi,
\eeq
where
\beq
\Phi = \frac1\varepsilon \sum_{ij}\int 
w_{ij}(x) P(x)\{{\cal E}(x^{ij}) - {\cal E}(x)\}  dx
\eeq
is the flux of ${\cal E}$ from the outside to the system.
It can be written as an average,
\beq
\Phi = \frac1\varepsilon \sum_{ij}
\la w_{ij}(x)\{{\cal E}(x^{ij}) - {\cal E}(x)\}\ra,
\eeq
and in the limit $\varepsilon\to0$, it becomes
\beq
\Phi = \sum_{ij} \la w_{ij}D_{ij}{\cal E}\ra.
\eeq
Here it is also convenient to write $\Phi$ as a sum of terms,
\beq
\Phi = \sum_{ij} \Phi_{ij},
\eeq
where
\beq
\Phi_{ij} = \la w_{ij}D_{ij}{\cal E}\ra,
\label{33}
\eeq
which is nonzero only when there is an arrow from $j$ to $i$.
In other cases $\Phi_{ij}$ vanishes.

\subsection{Relation between the entropy and energy fluxes}

To identify ${\cal E}$ with the energy function we 
identify $\Phi_{ij}$ as a heat flux. This is accomplished
by assuming the Clausius relation which says that 
the entropy flux is proportional to the heat flux,
the coefficient of proportionality being the inverse of
temperature. Accordingly, we assume that $\Phi_{ij}$
is related to $\Psi_{ij}$ by 
\beq
\Psi_{ij} = - \frac1T\Phi_{ij},
\label{35}
\eeq
where $T$ is understood as the temperature of the
ecosystem, identifying thus $\Phi_{ij}$ as the heat
flux and ${\cal E}$ with the energy function. 
We remark that (\ref{35}) is valid whenever $\Psi_{ij}$
is nonzero. 

To determined the actual expression of ${\cal E}(x)$,
we employ the condition (\ref{35}). 
Using the expressions (\ref{32}) and (\ref{33}),
a sufficient condition for (\ref{35}) is
\beq
D_{ij} w_{ij} = \beta w_{ij}D_{ij}{\cal E},
\eeq
where $\beta=1/kT$, which can be written as
\beq
D_{ij}\ln w_{ij} = \beta D_{ij}{\cal E}.
\eeq
Taking into account that $\ln w_{ij}$ is linear in
$\ln x_i$ and $\ln x_j$, then we wee that
${\cal E}$ can be chosen to be of the form
\beq
{\cal E} = E \sum_{i=0}^n \ln x_i,
\eeq
where $E$ is a positive constant.

Using this expression we determine the energy flux $\Phi_{ij}$
using formula (\ref{33}). For the first type of transition
rate, given by (\ref{31a}), we find
\beq
\Phi_{ij} = - b_{ij} E(s_i - s_j) \qquad i=1,2,\ldots,n.
\eeq
For the second type of transition rate, given by (\ref{31b}), we find
\beq
\Phi_{n+1,j} = - c_j E.
\eeq
For the third type of transition rate, given by (\ref{31c}), we find
\beq
\Phi_{0,n+1} = q E \la\frac1{x_0}\ra,
\label{37}
\eeq
which for small $\varepsilon$ gives the sult
as 
\beq
\Phi_{0,n+1} = \frac{q E}{s_0}.
\eeq
Comparing these three expressions with (\ref{33a}), (\ref{33b}),
and (\ref{33c}), we see that whenever $\Psi_{ij}$ is nonzero then
it is proportional to $\Phi_{ij}$, as long as $\beta=1/E$, or
$E=kT$, establishing the relation
between the entropy flux and the energy flux, understood as heat flux.

Let us defined $\Phi_i$ the flux of energy associated to $i$.
The total energy flux $\Phi$ is a sum of these
fluxes,
\beq
\Phi = \sum_i \Phi_i.
\eeq
In the stationary state $\Phi_i=0$ and $\Phi=0$. The flux of
energy from the environment to the plants is $\varphi=\Phi_{0,n+1}$,
which is the input of energy per unit time to the ecosystem and is
understood as the energy flux through the ecosystem,
and is given by (\ref{37}), that is,
\beq
\varphi = \frac{qE}{z_0}.
\eeq

Let us write $\Phi'$ as the sum of all nonzero fluxes $\Phi_{ij}$
except $\Phi_{0,n+1}=\varphi$,  
\beq
\Phi = \Phi' + \varphi.
\eeq
If we multiply $\Phi'$ by $-1/T$ we get $\Psi$,
that is
\beq
\Psi = - \frac1T \Phi'.
\eeq
Therefore
\beq
\beta\Phi = - \Psi + \beta \varphi.
\eeq
In the stationary state $\Phi=0$ and we reach the
result
\beq
\Psi = \frac1T \varphi >0.
\label{49}
\eeq
That is, the ecosystem is continuously producing entropy
which is throwing away at a rate equal to $ \varphi/T =  q E/T s_0$, or
considering that $E=kT$, it is 
\beq
\Psi = k \frac{q}{s_0}.
\label{50}
\eeq
We remark that the ratio $q/s_0$ can be understood as the rate at which
organic mass is produced by the plants per unit mass of the
plants.

\subsection{A simple model}

We apply the results above for the case of a simple
ecologic model consisting of three trophic levels,
each one consisting of just one species. The lower
level is represented by the plants ($i=0$),
the intermediate level by a herbivorous ($i=1$), and
the upper level by a carnivorous ($i=2$).
The nonzero transition rates are
\beq
w_{10}=ax_0x_1, \qquad w_{21}=bx_1x_2, \qquad w_{32}=cx_2,
\eeq
and $w_{30}=q$, where the state $i=3$ represent the environment.

The equations that give the time evolution of
the averages $s_i=\la x_i\ra$ are
\beq
\frac{ds_1}{dt} = as_0 s_1 - b s_1 s_2,
\eeq
\beq
\frac{ds_2}{dt} = b s_1 s_2 - c s_2,
\eeq
\beq
\frac{ds_0}{dt} = q - as_0s_1,
\eeq
\beq
\frac{ds_3}{dt} = - q + cs_2.
\eeq

In the stationary state
\beq
cs_2 - as_0s_1 = 0,
\eeq
\beq
as_0 s_1 - b s_1 s_2 = 0,
\eeq
\beq
b s_1 s_2 - c s_2 = 0,
\eeq
and $q = cs_2$.
Solving with the condition $s_0+s_1+s_2=1$, we get
\beq
s_1 = \frac{c}b, \qquad s_2 = \frac{a(b-c)}{b(a+b)},
\qquad s_0 = \frac{b-c}{a+b}.
\eeq

The energy function is given by
\beq
{\cal E} = E(\ln x_0 + \ln x_1 + \ln x_2),
\eeq
where we are setting $E=1$, so that $\beta=1$.
The energy fluxes at the stationary states are
\beq
\Phi_{10} = Ea(s_0-s_1) = E\frac{a}{b}\frac{b^2-2bc-ac}{a+b},
\eeq
\beq
\Phi_{21} = Eb(s_1 - s_2) = E\frac{2ac+bc-ab}{a+b},
\eeq
\beq
\Phi_{32} = - E c,
\eeq
\beq
\Phi_{03} = E\frac{q}{s_0} = E\frac{cs_2}{s_0} = E\frac{ac}{b},
\eeq
and the entropy fluxes are $\Psi_{10}=-\Phi_{10}/T$
$\Psi_{21}=-\Phi_{21}/T$, $\Psi_{32}=-\Phi_{32}/T$,
and $\Psi_{03}=0$. 
Notice that $\Phi=0$ and $\Psi=\Phi_{03}/T=kac/b$.

\section{Discussion and conclusion} 

We presented a stochastic approach to the population dynamics as
well as to the thermodynamics of ecosystems. The stochastic dynamics
leaded to a description in terms of a master equation which in
turn is approached by a Fokker-Planck equation. Within the present
stochastic approach, the fluctuations in the populations of the
several species are measured by a parameter $\varepsilon$ which is
inversely proportional to the total number size of the ecosystem. 

The dynamics are understood as a consequence of processes 
involving the several species that are analogous to chemical
reactions. The creation of an individual of a certain species is
analogous to an autocatalytic reaction. As a consequence if 
for some reason the number of individuals of a certain species
vanishes, then it becomes zero forever, that is, the species
becomes extinct. With the deterministic approach this occurs
for example when the feeding rate constant associated to 
an species is decreased. Observing figure \ref{traline},
if the parameter $b$ is decreased, the apex carnivorous is
extincted as one cross each one of the transition lines.

When $\varepsilon$ is small, which we consider to be always the case,
the equations for the main values of the number of individual
of each species coincides with the dynamic equations of the
deterministic approach. Thus we may understand the evolution
of a population as given by the deterministic equations supplemented
by the stochastic fluctuations. The fluctuations may lead to 
the extinction of a species, which is not the absolute type of
extinction explained above and predicted by the
deterministic approach. If by fluctuations the number of individuals
vanishes then the species becomes extinct, a type of
extinction that we call extinction by fluctuations.
If a small group of individuals of the extincted species is introduced
then the number of individuals may grow and the species is
restored. In the case of the absolute extinction this is not feasible,
that is, the small group of individuals will disappear instead
of growing.

The stochastic approach allowed us to formulate a nonequilibrium
thermodynamics theory for the ecosystem. This was acomplished
by means of two quantities: the entropy flux and heat flux.
The first was obtained from two important concepts which are
the entropy of the system, defined by the Gibbs expression
and the rate of entropy production. Here we used a definition
of entropy production which is appropriate for probability
transitions that do not have its reverse, that is, unidirectional
transitions. This definition leads us to an expression for
the entropy flux which can be understood as an average over
the probability distribution of the system. We then assume
that the flux of heat is proportional to the flux of entropy,
which is a relation conceived by Clausius. This allowed
us to obtain an expression for the energy function, 
establishing the thermodynamics of the ecosystem. 

A single main feature resulting from the present thermodynamic
approach concerning a food web consisting of plants,
animals and environment is as follows. The plants
transforms inorganic substances into organic matter
through photosynthesis inducing a flux of matter and
energy across the ecosystem. Simultaneously, entropy
is being generated and throwing away into the environment.
The entropy flux to the environment is given by
formula (\ref{50}), $\Psi=k (q/s_0)$, where $q/s_0$
is understood as the quantity of organic matter
generated by the plants per unit time per unit
mass of the plants. The flux of heat across the
ecosystem $\varphi$, according to equation (\ref{49}) is 
$\varphi=T\Psi$ where $T$ is understood as the temperature of
the ecosystem.



\begin{thebibliography}{99}

\bibitem{warming1909} E. Warming, {\it Oecology of Plants:
An Introduction to the Study of Plant Communities}
(Clarendon Press, Oxford, 1909).

\bibitem{tansley1935} A. G. Tansley,
Ecology {\bf 16}, 284 (1935).

\bibitem{elton1927} C. Elton, {\it Animal Ecology}
(Macmillan, New York, 1927).

\bibitem{odum1953} E. P. Odum {\it Fundamentals of Ecology},
(Saunders, Philadelphia, 1971).

\bibitem{macarthur1955} R. MacArthur,
Ecology {\bf 36}, 533 (1955).

\bibitem{odum1983} E. P. Odum, {\it Basic Ecology}
(CBS College Publishing, New York, 1983).

\bibitem{ricklefs2000} R. E. Ricklefs and G. L. Miller,
{\it Ecology} (Freeman, New York, 2000); 4th ed.

\bibitem{ricklefs2008} R. Ricklefs,
{\it The Economy of Nature}
(Freeman, New York, 2008); 6th ed.

\bibitem{paine1966} R. T. Paine,
American Naturalist {\bf 100}, 65 (1966).

\bibitem{may1973} R. M. May, {\it Stability and
Complexity in Model Ecosystems}
(Princeton University Press, Princeton, 1973).

\bibitem{smith1974} J. Maynard Smith, {\it Models in Ecology}
(Cambridge University Press, Cambridge, 1974).

\bibitem{pimm1982} S. L. Pimm, {\it Food Webs}
(Chapman and Hall, London, 1982).

\bibitem{deruiter2005} P. de Ruiter, V. Wolters, and J. C. Moore,
{\it Dynamic Food Webs: Multispecies, Assemblages, Ecosystem
development, and Environment Change}
(Elsevier, Amsterdam, 2005).

\bibitem{lotka1922} A. J. Lotka, 
Proc. Natl. Acad. Sci. {\bf 8} 147 (1922).

\bibitem{lotka1925} A. J. Lotka, {\it Elements of Physical
Biology} (Williams and Wilkins, Baltimore, 1925).

\bibitem{volterra1931} V. Volterra,
{\it Leçons sur la Théorie Mathématique de la
Lutte pour la Vie} (Gauthier-Villars, Paris, 1931).

\bibitem{may1971} R. May,
Mathematical Biosciences {\bf 12}, 59 (1971).

\bibitem{pimm1979} S. L. Pimm, 
Theoretical Population Biology {\bf 16}, 144 (1979).

\bibitem{pimm1987} S. L. Pimm and J. C. Rice,
Theoretical Population Biology {\bf 32}, 303 (1987).

\bibitem{moore1993} J. C. Moore, P. C.de Ruiter, and H. W. Hunt,
Science {\bf 261}, 906 (1993).

\bibitem{deruiter1995} P. C. de Ruiter, A.-M. Neutel, and 
J. C Moore,
Science {\bf 269}, 1257 (1995).

\bibitem{lindeman1942} R. L. Lindeman,
Ecology {\bf 23}, 399 (1942).

\bibitem{morowitz1968} H. J. Morowitz,
{\it Energy Flow in Biology},
Academic Press, New York, 1968.

\bibitem{dedonder1927}
T. De Donder, {\it L'Affinité}
(Gauthier-Villars, Paris, 1927).

\bibitem{prigogine1946} I. Prigogine et J. M. Wiame,
Experientia {\bf 2}, 451 (1946).

\bibitem{prigogine1947} I. Prigogine, {\it Étude
Thermodynamique des Phénomènes Irreversibles}
(Desoer, Liège, 1947).

\bibitem{prigogine1955} I. Prigogine, {\it Introduction
to Thermodynamics of Irreversible Processes}
(Thomas, Springfield, 1955).

\bibitem{glansdorff1971} P. Glansdorff and I. Prigogine,
{\it Thermodynamic theory of Structure, Stability and
Fluctuations} (Wiley, London, 1971). 

\bibitem{nicolis1977} G. Nicolis and I. Prigogine,
{\it Self-Organization in Nonequilibrium Systems}
(Wiley, New York, 1977).

\bibitem{zotin1967} A. I. Zotin and R. S. Zotina,
J. Theoret. Biol. {\bf 17}, 57 (1967).

\bibitem{feistel1981} R. Feistel and W. Ebeling,
Studia Biophysica {\bf 86}, 237 (1981).

\bibitem{assimacopoulos1986} D. Assimacopoulos,
Appl. Math. Modelling {\bf 10}, 234 (1986).

\bibitem{auger1989} P. Auger, {\it Dynamics and Thermodynamics
in Hierarchically Organized Systems}
(Pergamon Press, Oxford, 1989)

\bibitem{chakrabarti1995}
C. G. Chakrabarti, S. Ghosh, and S. Badhra,
Journal of Biological Physics {\bf 21}, 273 (1995).

\bibitem{svirezhev2000} Y.M.Svirezhev, 
Ecological Modelling {\bf 132}, 11 (2000).

\bibitem{jorgensen2004}
S. E. J{\o}rgensen and Y. M. Svirezhev,
{\it Towards a Thermodynamics Theory for Ecological Systems}
(Elsevier, Amstgerdam, 2004).

\bibitem{michaelian2005} K. Michaelians,
Journal of Theoretical Biology {\bf 237}, 323 (2005).

\bibitem{chakrabarti2009} C.G. Chakrabarti and K. Ghosh,
Ecological Modelling {\bf 220}, 1950 (2009).

\bibitem{andrae2010}
B. Andrae, J. Cremer, T. Reichenbach, E. Frey,
Phys. Rev. Lett. {\bf 104}, 218102 (2010).

\bibitem{nielsen2020}
S. N. Nielsen, F. Müller, J. C. Marques, S. Bastianoni, and 
S. V. J{\o}rgensen,
Entropy {\bf 22}, 820 (2020). 

\bibitem{tome2010} T Tomé, M. J. de Oliveira,
Phys. Rev. E {\bf 82}, 021120 (2010).

\bibitem{tome2015} T. Tomé and M. J. de Oliveira,
Phys. Rev. E {\bf 91}, 042140 (2015).

\bibitem{tome2018} T. Tomé and M. J. de Oliveira,
J. Chem. Phys. {\bf 148}, 224104 (2018).

\bibitem{tome2023a} T. Tomé, C. E. Fiore, and M. J. de Oliveira,
Phys. Rev. E {\bf 107}, 064135 (2023).

\bibitem{schnakenberg1976} J. Schnakenberg,
Rev. Mod. Phys. {\bf 48}, 571 (1976).

\bibitem{busiello2020} D. M. Busiello, D. Gupta, and A. Maritan,
Phys. Rev. Res. {\bf 2}, 023011 (2020).

\end{thebibliography}
\end{document}